\documentclass[12pt,a4paper]{article}
\usepackage{dfttrob}
\usepackage{amsmath}
\usepackage{latexuseful2e}
\usepackage{graphicx}
\newcommand{\citelow}[1]{\cite{#1}}
\newcommand{\abstracts}[1]{\begin{center}\textbf{Abstract}\end{center}
\begin{quote}#1\end{quote}}
\newcommand{\bFB}{b_{\text{FB}}}
\newcommand{\xx}[1]{\textit{#1}\hspace*{0.4em}}
\DeclareTextFontCommand{\zapf}{\fontencoding{U}\fontfamily{pzd}\selectfont}

\begin{document}

\dfttnum{DFTT 29/2002}

\title{FORWARD-BACKWARD MULTIPLICITY CORRELATIONS IN 
	\lowercase{\ee} ANNIHILATION
  AND \lowercase{pp} COLLISIONS AND THE WEIGHTED SUPERPOSITION 
	MECHANISM}

\author{A. GIOVANNINI AND R. UGOCCIONI\\
\small\itshape
Dipartimento di Fisica Teorica and I.N.F.N. - sezione di Torino \\
\small\itshape via P. Giuria 1, I-10125 Torino, Italy}

\maketitle

\abstracts{
Forward-backward multiplicity correlations in symmetric collisions are
calculated independently of the detailed form of the corresponding
multiplicity distribution.
Applications of these calculations to \ee\ annihilation and p\=p
collisions confirm the existence of the weighted superposition
mechanism of different classes of substructures or components.
When applied to p\=p collisions in particular, clan concept and its
particle leakage from one hemisphere to the opposite one become of
fundamental importance.
The increase with c.m.\ energy of the correlation strength as well as
the behaviour of the average number of backward particles vs.\ the
number of forward particles are correctly reproduced.
}

\vspace{2cm}
\begin{center}
Talk presented by A. Giovannini at the
X International Workshop on Multiparticle Production
``Correlations and Fluctuations 2002'', 
Crete, Greece, 8-15 June 2002.
\end{center}
\newpage

\section{Essentials on forward-backward multiplicity  correlation in
symmetric collisions}

The  average number of charged particles generated in different
events  in the backward hemisphere (B), $\nbar_B$, is a function 
of the number of particles occurring  in the forward hemisphere (F),
$n_F$, 
controlled by  the correlation strength  $\bFB$
\begin{equation}
 \bFB = \frac{\text{Cov} [ n_B,n_F ] }{
			\sqrt{ \text{Var} [n_B] \text{Var} [n_F] } } . \label{eq:1}
 \end{equation}

In hadron-hadron collisions\cite{UA5:FB,ISR:FB,NA22:FB,FB:E735}
the correlation  strength parameter
is rather large with respect to \ee\ annihilation\cite{OPAL:FB,Tasso:FB}
and is
growing with c.m.\ energy in the total sample of events as shown in
Table \ref{tab:1}.

\begin{table}[b]
  \caption{Experimental results on forward-backward correlation strength.}\label{tab:1}
	\begin{center}
  \begin{tabular}{|rrll|}
	\hline
		 &     & \multicolumn{1}{c}{$\bFB$}& \\
	\hline
	p\=p & UA5 & $0.43 \pm 0.01$ ($1 < |\eta| < 4$)  & 546 GeV c.m.\
	energy\\
	   &     & $0.58 \pm 0.01$ ($0 < |\eta| < 4$) & \\
	pp & ISR & $0.155 \pm 0.013 $ & 63 GeV c.m.\ energy\\
	\hline
	\ee & OPAL & $0.103 \pm 0.007$  & LEP \\
			& TASSO & $0.080 \pm 0.016$ & 22 GeV c.m.\ energy\\
	\hline
  \end{tabular}
	\end{center}
  \end{table}

In addition in \ee\ annihilation at LEP energies it has been
found\cite{OPAL:FB}
that $\bFB \approx 0$ in the separate  two- and three-jet sample
of events.
No  information is available on  the correlation strength
in the separate samples of soft (no minijets) and semihard (with minijets)
events in hadron-hadron collisions.

\section{The problem}

We want to calculate the parameter $\bFB$  for the multiplicity distribution
\begin{equation}
P(n) =  \sum_{n_B+n_F = n}   P_{\text{total}} (n_F,n_B) , \label{eq:2}
\end{equation}
where $n_F$ and $n_B$ are random variables and $P_{\text{total}}
(n_F,n_B)$  is the joint 
distribution for the weighted superposition of different classes of 
events,\cite{combo:prd} i.e.,
\begin{equation}
P_{\text{total}} (n_F,n_B) = \alpha P_1 (n_F,n_B) + 
		( 1 - \alpha )  P_2 (n_F,n_B) ,
		\label{eq:3}
\end{equation}
$\alpha$ being the weight of class 1 events with respect to the total.

\section{The general solution}
\begin{equation}
	\bFB = \frac{
		\begin{array}{cc}\alpha b_1 {D^2_{n,1}}(1+b_2) +
				(1-\alpha) b_2 {D^2_{n,2}}(1+b_1) +\qquad\qquad\qquad\qquad\\
					\qquad\qquad\qquad\qquad+\frac{1}{2}\alpha(1-\alpha)(\nbar_{2} -
       \nbar_{1})^2(1+b_1)(1+b_2)
		\end{array}}
			{\begin{array}{cc}\alpha  {D^2_{n,1}}(1+b_2) +
				(1-\alpha)  {D^2_{n,2}}(1+b_1) +\qquad\qquad\qquad\qquad\\
					\qquad\qquad\qquad\qquad+\frac{1}{2}\alpha(1-\alpha)(\nbar_{2} -
        \nbar_{1})^2(1+b_1)(1+b_2)\end{array}}  ,
																				\label{eq:b_total}
\end{equation}
%\bFB = ....    FORMULA 20 del lavoro                          [A]
where $b_i$ are the correlation strengths of class  1  ($i=1$) and
class 2 ($i=2$) events, $D_{n,i}$ are the multiplicity distribution dispersions
of class 1 ($i=1$) and class 2  ($i =2$) events and $\nbar_i$ the 
corresponding average charged multiplicity  for class 1 ($i=1$) and
class 2 ($i=2$) events.

In case $b_1$ and $b_2$ are zero (as in the separate two samples of events
in \ee\ annihilation) one finds
\begin{equation}
	\bFB = \frac{
					\frac{1}{2}\alpha(1-\alpha)(\nbar_{2} - \nbar_{1})^2}
			{\alpha  {D^2_{n,1}} +
				(1-\alpha)  {D^2_{n,2}} +
					\frac{1}{2}\alpha(1-\alpha)(\nbar_{2} - \nbar_{1})^2} .
																					\label{eq:b_12}
\end{equation}

It should be pointed out that above formulas are independent from any specific
form of the multiplicity distributions $P_1$ and $P_2$!
They depend only on the weight alpha and average charged multiplicities
and dispersions of the two classes of events.

\section{Applications of Eqs.~(\ref{eq:b_total}) and (\ref{eq:b_12})}

\subsection{An intriguing application of Eq.~(\ref{eq:b_12})  
	to \ee\ annihilation}

Opal collaboration has found that forward backward multiplicity correlations
are non existent in the separate two- and three-jet samples of events
i.e. $b_1$ and $b_2$ in the first general formula are zero and the correlation
strength of the total sample of 2-jet and 3-jet  events is equal 
to $0.103 \pm 0.007$.

Using a fit to OPAL data with similar conditions to the jet finder
algorithm for the separate samples of events we can determine all
parameters in formula (\ref{eq:b_12}) and test its prediction with the
experimental finding.

It turns out that  the values of the parameters\cite{hqlett:2} needed in 
(\ref{eq:b_12}) are $\alpha=0.463$,
$\nbar_1=18.4$, $\nbar_2 = 24.0$, $D^2_1=25.6$, $D^2_2=44.6$ and
the predicted value of $\bFB$ is  0.101, in extraordinary agreement with
experimental data!

\subsection{A  suggestive application of Eq.~(\ref{eq:b_total}) 
	to p\=p collisions}

The application of (\ref{eq:b_12}) to p\=p collisions leads to unsatisfactory results
but opens a new perspective: forward-backward multiplicity correlations
cannot be neglected in the separate components.
Accordingly Equation (\ref{eq:b_total}) and not
(\ref{eq:b_12}) should be used.

Repeating the same approach  done  in \ee\ annihilation for calculating 
$\bFB$ (i.e., assuming that in the separate samples of events FB multiplicity
correlations are absent, $b_1=b_2=0$) in the case of  p\=p collisions
at 546 GeV c.m.\ energy  and using Fuglesang's fit\cite{Fug}
to soft and semihard events  
(accordingly  $\alpha=0.75$, $\nbar_1=24.0$, $\nbar_2=47.6$, $D^2_{n,1} =106$,
$D^2_{n,2}=209$) one finds $\bFB=0.28$  ($\bFB^{\text{(exp)}}= 0.58$).
The theoretical prediction in this case is  too small! It is clear that
our working hypothesis was not correct in this case.
In conclusion forward-backward multiplicity correlations are needed in
each class of events, i.e., $b_1$ and  $b_2$ should be different from zero,
and after their determination    general formula (\ref{eq:b_total}) 
and not formula (\ref{eq:b_12})
should be used!

Results in 4.1 and 4.2 are a striking test of the existence of the
weighted superposition effect, only a guess up to now.

\section{A new theoretical problem }

Following above conclusions the next problem is 
how to determine $b_1$ and $b_2$  when explicit data on forward-backward 
multiplicity correlations in the two separate samples of events are lacking  
and $\bFB$ of the total sample is known from experiments.
 
The generality of Equation (\ref{eq:b_total}) 
should be limited by introducing additive 
assumptions inspired by  our phenomenological knowledge of  the particle 
emission process  in the collision under examination.

Assuming for instance that

\xx{a.}particles are independently produced in the collision,

\xx{b.}binomially distributed in the forward and backward hemispheres,

it is found that
\begin{equation}
b_i = \frac{D^2_{n,i} - \nbar_i}{D^2_{n,i} + \nbar_i}  ,
		\label{eq:4}
\end{equation}
where  $D_{n,i}$ and $\nbar_i$ are the dispersion and the average charged
multiplicity of the overall multiplicity distribution of each component
being as usual $i=1,2$.

Assuming next that 

\xx{c.}the multiplicity distribution in each $i$-component is  NB(Pascal)   
with parameters $\nbar_i$ and $k_i$ (an assumption which is suggested by the 
success of the weighted superposition mechanism of NB(Pascal)MD's  in describing shoulder effect in
charged particle multiplicity distributions and $H_q$ vs $q$ oscillations and
which we hardly would like to abandon),

we find
\begin{equation}
b_i = \frac{ \nbar_i }{ \nbar_i + 2  k_i } . \label{eq:5}
\end{equation}

Accordingly $b_i$ values can be calculated by using again  Fuglesang's fit
parameters on the two components at 546 GeV c.m.\ energy. After inserting 
in the general formula (\ref{eq:b_total}) these parameters we find
$\bFB= 0.78$. 

A too large value with respect to the experimental one ($\bFB = 0.58$)!
This result leads to the following question: Which  one of above mentioned
apparently quite reasonable  assumptions should be modified?

Our guess  is that charged particle FB multiplicity correlation  is not
compatible with independent particle emission but is compatible
with the production in cluster, i.e., clan within a NB(Pascal)MD
framework. An idea which   we propose to develop and to explore in
the following.

\section{Clan concept is of fundamental importance}

Successive  steps of  our argument\cite{RU:FB} are

i) the joint distribution $P_{\text{total}} (n_F,n_B)$ 
is written as the convolution
over the number of produced clans and over the partitions of forward
and backward produced particles among clans:
\begin{equation}
	P(n_F,n_B) = \sum_{N_F,N_B} {\cal P}(N_F,N_B) \!\!
				\mathop{\sum_{m_F'+m_F'' = n_F}}_{m_B'+m_B'' = n_B} \!\!
				p_F(m_F',m_B'|N_F) p_B(m_F'',m_B''|N_B) .
																				\label{eq:6}
\end{equation}

ii) forward backward hemispheres   symmetry  property  is used
\begin{equation}
	p_F(n,m|N) = p_B(m,n|N) .  \label{eq:7}
\end{equation}

iii) leakage parameter $p$ is introduced: it controls the probability
that a binomially distributed particle generated by one clan lying in
one hemisphere has to leak in the opposite hemisphere, $q$ is the
leakage parameter working in the symmetric direction, $p+q=1$ (notice
that $p=1$ or $q=0$ means no leakage, the variation domain of $p$ is
$0.5 \leq p < 1$ and when $p < 0.5$ the clan is classified in the
wrong domain).

iv) covariance $\gamma \equiv \avg{ (\mu_{F} - \bar\mu_{F}) 
(\mu_{B} - \bar\mu_{B})}$
of  $\mu_F$  forward and $\mu_B$  backward particles within a clan  for
forward and backward binomially distributed particles generated by clans
is also introduced.

v) clans are binomially produced in the forward and backward hemispheres
with the same probability and particles within  a clan are independently
distributed in the two hemispheres.

It follows for each $i$-component
\begin{equation}
	\begin{split}
	b &= \frac{D^2_N - 4\avg{d^2_{N_F}(N)}(p-q)^2 + 4\Nbar\gamma/\nc^2}
			{D^2_N + 4\avg{d^2_{N_F}(N)}(p-q)^2 - 4\Nbar\gamma/\nc^2 + 
					2\Nbar D^2_c/\nc^2}\\
		&= \frac{D^2_n/\nbar - D^2_c/\nc -
	4\avg{d^2_{N_F}(N)}(p-q)^2\nc/\Nbar + 4\gamma/\nc }{
	D^2_n/\nbar  + D^2_c/\nc +
	4\avg{d^2_{N_F}(N)}(p-q)^2\nc/\Nbar - 4\gamma/\nc }  .
	\end{split}
			\label{eq:C}
\end{equation}

Eq.~(\ref{eq:C}) assuming NB (Pascal) behavior with characteristic
$\nbar_i$ and $k_i$ parameters for each component, binomial clan
distribution in the two hemispheres, binomial distribution in the two
hemispheres of logarithmically produced particles from each clan
according to clan structure analysis gives
\begin{equation}
	b_i = \frac{ 2 \nbar_i p_i q_i }{\nbar_i + k_i - 2 \nbar_i p_i q_i } .
					\label{eq:Cstar}
\end{equation}

Accordingly the  problem is therefore  reduced  to determine leakage 
parameters $p_i$ in the two classes of events! 

Notice that in the limit $\nbar_i \to \infty$, for decreasing  $k_i$,
$b_i$ depends on $p_i$ only.

\section{A phenomenological  argument for determining leakage parameters $p_i$}

By assuming that the semihard component is negligible at 63 GeV c.m.\ 
energy and knowing $\bFB$ from experiment at such energy,
equation~(\ref{eq:Cstar}) 
allows to determine $p_{\text{soft}}$ (0.78); the relatively  small variation 
of $\nbar_{c,\text{soft}}$ from 63 GeV to 900 GeV 
(it goes from  $\approx 2$ to $\approx  2.44$)
leads to the conclusion that the leakage parameter  for the soft
component $p_{\text{soft}}$ can be considered in the GeV domain nearly constant,
i.e., $p_{\text{soft}}= 0.78$: therefore the correlation strength for the soft
component at 546 GeV c.m.\ energy, $b_{\text{soft}}(546~\text{GeV})$,  
can easily be determined.

The germane equation for $b_{\text{semihard}}(546~\text{GeV})$ 
contains of course the 
unknown parameter $p_{\text{semihard}}$ at the c.m.\ energy  of 546
GeV. By inserting in
equation (\ref{eq:b_total}) for $\bFB$ (total)  
$b_{\text{soft}}(546~\text{GeV}) =0.78 $ 
and $b_{\text{semihard}}(546~\text{GeV})$ 
as given by equation (\ref{eq:Cstar}) with unknown
$p_{\text{semihard}}$  parameter, $p_{\text{semihard}}$  at
546 GeV can be   calculated from the
experimental value of  $\bFB$ (total) = 0.58. 
It is found $p_{\text{semihard}}(546~\text{GeV})=0.77$.

Since $\nbar_{c,\text{semihard}}$ does not vary too much in the GeV region 
(it goes from 1.64 at 200 GeV c.m.\ energy to 2.63 at 900 GeV c.m.\
energy, a relatively small variation which will hardly affect the
corresponding leakage parameter in this domain) it is not hazardous
to take $p_{\text{semihard}} \approx$ constant in the same  region.

Under just mentioned assumptions

\begin{figure}
  \begin{center}
  \mbox{\includegraphics{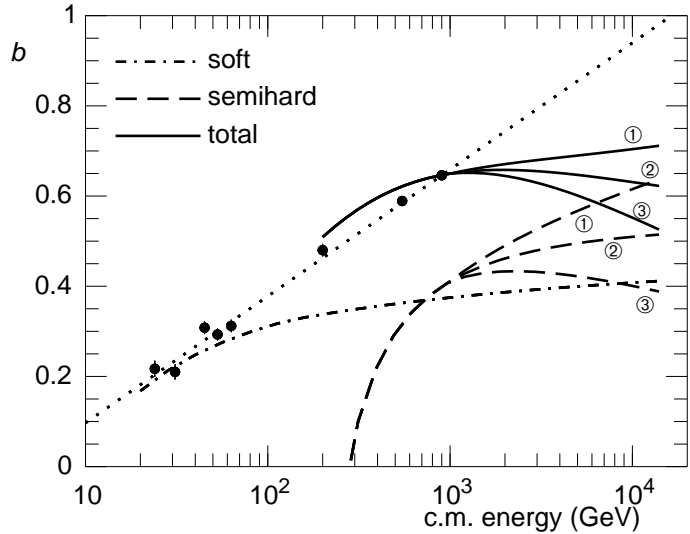}}
  \end{center}
  \caption{Predictions for the correlation coefficients 
	for each component (soft and semi-hard) 
	and for the total distribution in $p\bar p$ collisions
	in scenario 2. Three
	cases are illustrated, corresponding to the three numbered branches: 
  leakage increasing with $\sqrt{s}$ (upper branch, \zapf{\char'300}),
	constant leakage (middle branch, \zapf{\char'301}) and
	leakage decreasing with $\sqrt{s}$ (lower branch, \zapf{\char'302}).
	Leakage for the soft component is assumed constant at all energies.
	The dotted line is a fit to 
	experimental values.}\label{fig:1}
  \end{figure}

\xx{a.}the correlation strength c.m.\ energy dependence is correctly
reproduced in the GeV energy range from ISR up to UA5 top  c.m.\
energy and follows   the phenomenological formula $\bFB = - 0.019 +
0.061 \ln s$ (see Fig.~\ref{fig:1}).

\xx{b.}when extrapolated to the TeV energy domain in the scenarios discussed 
in Ref.~\citelow{combo:prd}  
with the same values of $p_{\text{soft}}$ obtained in
the GeV region 
($\nbar_{c,\text{soft}}(14~\text{TeV})$ being $\approx 2.98$ 
makes this guess acceptable)
and $p_{\text{semihard}}$  also constant (a too strong assumption of course),
a clean bending effect in $\bFB$ vs.\ $\ln s$  is predicted. Bending effect is
enhanced or reduced  by allowing $p_{\text{semihard}}$ to increase
(less leakage from 
clans and more bending) or to decrease logarithmically with c.m.\ energy
 (more leakage from clans and  less bending).
Energy dependence  of leakage parameter for the semihard component  is clearly
expected in the TeV region  in a scenario with strong KNO scaling violation 
in view of the quite large average number of  particles per clan 
with respect to that found at 900 GeV c.m.\ energy (it goes 
from 2.63 at 900 GeV  up to 7.36 at 14 TeV). See again Fig.~\ref{fig:1}.

\begin{figure}
  \begin{center}
  \mbox{\includegraphics{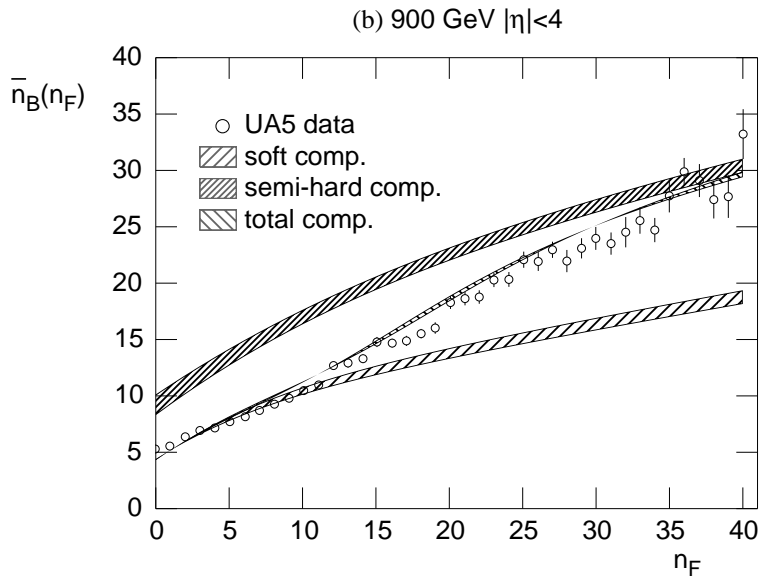}}
  \end{center}
  \caption{Results of our model for $\nbar_B(n_F)$ vs.\ $n_F$ compared 
  to experimental data in the
  pseudo-rapidity interval $|\eta|<4$ at 900 GeV.}\label{fig:2}
  \end{figure}

\xx{c.}in addition $\nbar_B (n_F)$  behavior at 63 GeV c.m.\ energy (ISR data)
is quite well described in terms of the soft component (single NB) only
and at 900 GeV c.m.\ energy  (UA5 data) in terms of the 
weighted superposition of soft and semihard components, i.e., of the
superposition of two NB(Pascal)MD's.  (See Fig.~\ref{fig:2}, where the
second case is shown).

\section{Conclusions}
Weighted superposition mechanism of two samples of events  describes
forward backward multiplicity correlations in \ee\
annihilation independently of the specific form of the charged particle
MD  in the different classes of events: only the  average numbers of 
particles and related dispersions in addition to the weight factor are needed.

In order to describe forward backward multiplicity correlations  in
p\=p collisions  lack of information on FB multiplicity correlations
in the separate components is demanding to specify  the form of particle
multiplicity distributions of the two components.

The choice of NB(P)MD for each component
(supported by its success in describing shoulder effect and $H_q$ vs $q$
oscillations)  outlines the role
of clan properties in this framework  and allows to determine
correctly $\bFB$ energy dependence for the total sample of events in the 
GeV region. Its bending in the TeV region within possible scenarios
discussed in the literature  is predicted.

$\nbar_B(n_F)$ vs $n_F$ trend is also nicely  reproduced at 63 GeV (only soft 
component is assumed to contribute)  and at 900 GeV (superposition of soft 
and semihard components is used), and its behavior in the TeV energy range
predicted.

Last but not least we have found that our study on FB multiplicity
correlations  in pp collisions when extended to the TeV energy region
assuming KNO scaling violation for the semihard component enhances
the intriguing connection already shyly   anticipated in the GeV 
region  between particle populations within clans, particle leakage
from clans in one hemisphere to the opposite one and the superposition
effect between different components. Clan concept appears in this
framework as a powerful tool which goes far beyond  its simple
statistical interpretation and raises the question on its real physical
significance: an interesting but also compulsory question for future
experimental work in pp collisions and not only.

% here is the bibliography
\section*{References}
\bibliographystyle{prstyR}
\bibliography{abbrevs,bibliography}

\end{document}